# Resilience Aspects in Distributed Wireless Electroencephalographic Sampling

R. Natarov[1], O. Sudakov[3,4], Z. Dyka[1], I. Kabin[1], O. Maksymyuk[4], O. Iegorova[4], O. Krishtal[4] and P. Langendörfer[1,2]

[1]IHP – Leibniz-Institut für innovative Mikroelektronik
Frankfurt (Oder), Germany

[2]BTU Cottbus-Senftenberg
Cottbus, Germany

[3]Medical Radiophysics Department
Taras Shevchenko National University of Kyiv
Kyiv, Ukraine

[4]Department of Cellular Membranology
Bogomoletz Institute of Physiology
Kyiv, Ukraine

*Abstract*—Resilience aspects of remote electroencephalography sampling are considered. The possibility to use motion sensors data and measurement of industrial power network interference for detection of failed sampling channels is demonstrated. No significant correlation between signals of failed channels and motion sensors data is shown. Level of 50 Hz spectral component from failed channels significantly differs from level of 50 Hz component of normally operating channel. Conclusions about application of these results for increasing resilience of electroencephalography sampling is made.

## I. Introduction

### A. Motivation

Corresponding to the U.S. National Institute of Standards and Technology (NIST) the resilience of Cyber Physical Systems (CPS) is the ability to anticipate different adverse and/or dangerous conditions [1]. Corresponding to this definition a resilient system has to be able to resist different conditions while providing the acceptable level of operation and services quality. System resilience is important for applications in different areas including telecommunications, logistics, transport, etc. Thus a large number of studies for resilience problems are available [2], [3]. Resilient technical systems have to be at least self-aware, self-adaptable and self-reconfigurable to be able to self-repair. Implementing these features is a complex task and can require the use of Artificial Intelligence (AI) algorithms. To be self-adaptable and self-reconfigurable the CPSs have to be able "to think" and decide. In our opinion, resilience is more than the sum of security and dependability i.e. cognition is what makes the difference [4].

Systems for monitoring, prediction and control of physiological parameters are widely used for clinical and research applications. Such systems are commercially available and under development. Distributed physiological monitoring systems include sensors [5], [6], massive storage and processing elements [7]-[9], algorithms and software for data analysis [10]-[12]. Standards for physiological monitoring sensors sets (body area network, BAN [13], [14]) are under development and implementation now. Resilience of such systems is of great importance [5]. Engineering of the resilient systems include a large variety of physical, technical, security and other issues whereby no universal solutions for providing system resilience exist. Systems for physiological monitoring have a variety of application specific resilience issues. These issues are related to biological and physical specificity of data sampling.

Epilepsy is a chronic condition of the brain triggered by a heterogeneous group of neurological disorders with diverse etiologies, behavioural seizure patterns and pharmacological sensitivities. This condition is responsible for a high level of suffering. It affects more than 50 million people worldwide representing an important public health problem. According to data of the European Forum on Epilepsy Research [15], about 6 million European citizens have epilepsy-related disorders and there are ~400,000 new cases in Europe each year, meaning one new case every minute. Unfortunately, up to 40% of patients suffer from epileptic attacks that cannot be adequately controlled by conventional pharmacotherapy. So, new means to control or at least predict new seizures are urgently needed. Here body area networks recording physiological parameters of the patients are a promising approach.

The use of wireless devices as a means to monitor physiological parameters such as pulse, movements etc. is becoming widely accepted and has found its way even in the consumer devices such as the Fitbit [16], etc. But monitoring more complex parameters and events as well as their analysis under real time conditions in order to improve numerous patients' lives is still in its infancy. It is known that sometimes epileptic patients can predict their seizures several minutes before the onset of them. This phenomenon is called aura, which is in fact a focal aware seizure [17]. Moreover the specially trained dogs can do the same [18]. If a special device that can predict the epileptic seizure with a high accuracy will be designed, patients can take precaution actions to avoid severe consequences or even supress seizure onset with appropriate medications. Research in the field of epilepsy requires sampling of large amount of physiological experimental data from patients and laboratory animals. The prediction of epileptic seizures demands complicated data analysis algorithms, e.g. artificial intelligence or numerical simulations. The design of smart resilient sensors for such data sampling is of great interest.

*B. Resilience in EEG sampling*

Activity of thousands or millions of neurons in the brain generates electrical potential fluctuations resulting in measurable EEG signals. EEG signals in the "normal" state significantly differ from the ones during an epileptic seizure. As it was mentioned before, seizures are often preceded by aura [17]. One of the challenges here is the short time between first signs of the epileptic seizure and its occurrence requiring extremely fast data transfer and processing.

Useful EEG information is located in low spectral frequencies from about 0 to 60 Hz and sometimes to several hundred Hertz. The sampling rate applied to capture the EEG is usually in the range of 256 Hz (i.e. 256 Samples/sec) to several thousand Hz. The potential fluctuations can be sampled using non-invasive electrodes placed on the head skin (scalp) or intracranial (implanted) electrodes.

The common number of electrodes for intracranial measurement varies from one to thousands [19]. Intracranial electrodes conduct signal values up to hundred millivolts with noise level in the range of microvolts.

Measurements of electric potentials using non-implanted (contacted) electrodes on the scalp are more applicable to humans. Fig. 1 shows some examples of such non-invasive electrodes. The measured values are in a range from several microvolts to several hundred microvolts with a noise level up to several microvolts or even higher i.e. the signal/noise ratio is 100 times less than in case of implanted electrodes [20]. Some examples of contact electrodes are shown in Fig. 1: most common are wet ones, see *(a), (b)*; they are used with a conductive gel to achieve low impedance, but their preparation for measurements is time-consuming. Dry electrodes, see Fig. 1-*(c)*, are much easier to use and their contact quality is comparable to the one of the wet ones [21].

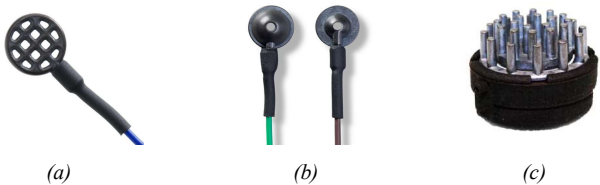

Figure 1. Examples of the commercial contact electrodes for EEG measurements: *(a)* Webb and *(b)* Cup electrodes [22], *(c)* Wearable Sensing (one electrode) [23]

Common contact EEG uses from 10 to several hundreds of sampling channels. EEG signals have to be measured, amplified, transmitted/received and stored for further processing. EEG data transmission is preferred to be wireless due to patient's movement during trials. Real time EEG data processing should implement detection and filtration of such movements. Due to the low EEG signal amplitude measured using non-implanted electrodes, contact EEG is very sensitive to:

- different kinds of human motions;
- contact stability;
- interference from industrial power supply network (the range of 50-60 Hz falls into the EEG spectral range).

Notch filters and screening are used to mitigate interference, nevertheless it is always present in real life signals. Motion of patients is a main source of signal artefacts. Most important are myogram (signal from body muscles) and oculogram (signals from eyes motion). Electric potentials induced by muscles significantly exceed the EEG signals and affect the information of the EEG. Moreover, the motion of patients or laboratory animals and drifting of the amplifier's parameters lead to a change in the electrode contact resistance, the increase of noise or the degradation of the signal.

The poor contact quality should be detected and fixed as soon as possible and this detection and correction should be performed at the sampling device side. Special actions need to be taken to increase contact quality [10], nevertheless long time monitoring is sensitive to such issues. A large number of papers are related to detection of failed EEG channels (e.g. [24]-[26]) as well as to detection and correction of EEG artefacts (e.g. [8], [26]). Most of them use only the measured EEG signals for the analysis, whereby the analysis can include the linear approaches (wavelets, Fourier analysis, correlation analysis), artificial intelligence approaches (support vector machines, artificial neuronal networks), etc.

In this work we concentrate on the detection of failed/poor contacts and investigate the possibility to use additional sensors of wearable EEG device to do so. These additional sensors can be motion detectors e.g. accelerometer and/or body position sensors e.g. magnetometer. Our investigations are based on the following assumptions:

- the signal range from electrodes with a poor contact are low and spikes from bad contact are present;
- the signals from electrodes with poor contact should have correlation with motion due to changes of the contact resistance.

Additionally we investigated the possibility to also take interference signals from industrial power network into account. Combining the correlation of EEG signals with signals from motion sensors and with signals from the industrial power network can improve the detection of the poor contact artefact in EEG traces.

The rest of this paper is structured as follows. In section II we describe our experimental setup. In sections III and IV we explain how we processed the data and discuss the results of our analysis. Conclusions are given in section V.

## II. EXPERIMENTAL SETUP

For the recording and further analysis of EEG data we used the low-cost general-purpose commercial headset Emotiv Epoc Plus (2018 year version). Key features of this non-medical EEG acquisition system are:

- 16 EEG channels (14 in operation, 2 as reference),
- 256 Hz sampling rate with 16 bit resolution,
- embedded accelerometer and magnetometer with 64 Hz sampling rate,
- built-in rechargeable battery,
- wireless communication,
- saline soaked contact pads with simplified positioning method [27].

This device shows better performance than similar solutions used in research [28]. Data from the headset are transferred via 2.4 GHz radio channel with proprietary protocol to the computer with USB receiver. Additionally it is possible to grab the data via Bluetooth Low Energy (BLE) with an application for Android/iOS smartphones provided by the manufacturer. A general connection diagram for the device in our experiments is shown in Fig. 2.

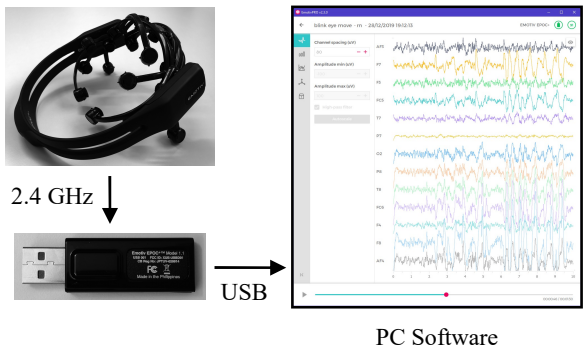

2.4 GHz

USB

PC Software

Figure 2. Our measurement setup: portable wireless EEG headset

The manufacturer's software allows to monitor EEG and auxiliary data in real time, provides spectral EEG signal representation and specific characteristics (excitement, fear, stress, happiness, etc.), the battery level, external marks and events from the COM port or the USB input device.

We selected the Emotiv Epoc Plus headset due to the following features:

- wireless connection grants some mobility to a test subject and increases safety as high voltage discharge through cables is impossible;
- data export to popular formats;

- visual representation of the pads' connection quality in real time that we used for the evaluation of our experiments.

From our point of view the headset has some drawbacks:

- Due to the online authorization of the provided software the headset works only with an operational internet connection. This limits its use particularly when the internet connection is limited or absent.
- Occasional drops and data stuttering between the headset and the USB receiver at a distance less than 3 meters. The headset does not contain any internal memory thus the data measured during these drops are lost.
- No audio/visual indication about low battery and out of range position.
- Some persons state that continuous wearing of the headset for longer than 1 hour leads to headache, nevertheless they recovered within short time after the headset was putted off.

Here we analysed the data measured from a healthy person during different kinds of movements: walking, head movements/shaking and eyes blinking. These data were collected from 14 EEG channels, 3 accelerometer channels for X, Y, Z axes, and 3 magnetometer channels for X, Y, Z axes.

Fig. 3 shows schematically the location of the EEG electrodes on a human head and their contact quality. The headset performs conductivity measurements to derive the contact quality of the electrodes [29]. This metrics is represented by integer values from 0 (low contact quality, marked in red) to 4 (good contact quality, marked in green), updated in real time during the measurements and indicated in the software. Also contact quality measurements can be exported, and we used it to evaluate our analysis results.

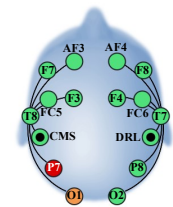

Figure 3. Location of the electrodes on a head and colour-coded contact quality representation during measurements

We recorded the data for about 60 seconds during walking and head motion and 24 seconds for eyes blinking. During the test no data drops were detected.

## III. PROCESSING COLLECTED DATA

Examples of signals sampled during walking, head motion and eyes blinking are shown in Fig. 4.

Electrodes P7 and O1 a have poor contacts during all of our experiments (see lines marked red and orange in Fig. 4).

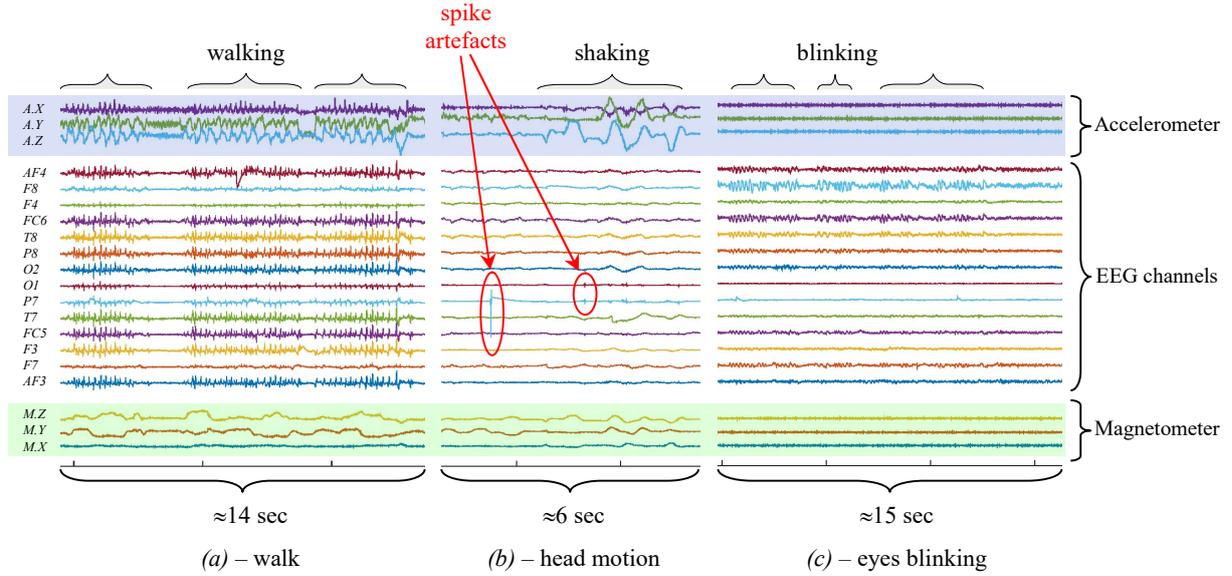

Figure 4. Signals sampled during walking *(a)*, head motion *(b)* and eyes blinking *(c)*. The 3 top lines represented on the light blue background are signals from 3 accelerometer channels. The 3 bottom lines (green background) are signals from 3 magnetometer channels of the EEG headset. 14 lines in the middle (white background) are EEG signals

Also some channels show waveforms that differ from other channels during different tests:

- signals from the electrodes F8, F4 and F7 are weak during walking, see Fig. 4-*(a)*;
- signals from the electrodes F3 and AF3 are sometimes weak during head shaking, see the first 3 seconds in Fig. 4-*(b)*;
- P7 presents spike artefacts, see Fig. 4-*(b)*.

Additionally, Fig. 4 illustrates the following facts:

- Walking is characterised by moderate accelerometer signals and strong magnetometer signals.
- Head motion produces strong accelerometer and moderate magnetometer signals.
- Eyes blinking generates no accelerometer and magnetometer signals, only noise can be seen in Fig. 4-*(c)*.

The measured signals for each kind of motion were represented as a separate matrix $S$ that contains different channels in columns: 14 columns are EEG signals, 3 columns are signals measured from magnetometer and 3 columns are accelerometer signals. Taking into account that EEG signals were sampled at 256 samples/second and motion detectors signals were sampled at 64 Hz rates the EEG signals in matrices S were reduced to 64 samples/second. Each channel was normalised by subtraction of its mean value and scaled by its standard deviation.

We applied the Principal Components Analysis (PCA) for investigating the correlation between motion sensor signals and EEG signals. We used the *svd* function in GNU Octave [30] to perform the PCA analysis. Singular values decomposition of the matrices $S$ for each motion type was performed:

$$S = U\Sigma V^T. \qquad (1)$$

Matrices $U, V$ are orthogonal and matrix $\Sigma$ is diagonal with positive values sorted in ascending order and equal to variances of principal components. Trace of matrix $\Sigma$ i.e. the sum of elements on the main diagonal of $\Sigma$ is the total variance. Columns of matrix $T = U\Sigma$ contain the principal components that are also called *scores*. Columns of matrix $V$ are called *loadings* and contain contributions of original signals to principal components.

The variance of the largest principal component PC1 for walk, head motion and eyes blinking does not exceed 20% of the total variance for each motion type. The smallest variance of PC20 is not smaller than 2% of the total variance. It means that no significant correlation between all signal channels exists and no signal channels may be determined from other signals channels by linear combination for all investigated motion types.

Correlation coefficients between individual channels are described by loadings matrix $V$. Channels that have large contributions to the same principal component are correlated or anti-correlated depending on the signs of their contributions.

Principal components PC1 and PC2 were selected for further analysis due to maximal informational impact. Loading plots of selected components for the walking and the head motion are shown in Fig. 5-*(a)* and Fig. 6-*(a)* respectively. Blue dots on each plot represent EEG channels, green – accelerometer, red – magnetometer.

To assess the effects of movements on EEG signals and to determine other dependences in our data, correlation between each feature (channels in our case) within PC1 and PC2 components were calculated. This was performed by representing each channel in PC1, PC2 coordinates as a vector (see Fig. 5-*(a)* and Fig. 6-*(a)*) and further computing the angle between each vector. The cosine of that angle is a desired correlation of the selected channels. Correlation maps for the walking and the head shaking are shown in Fig. 5-*(b)* and Fig. 6-*(b)* respectively.

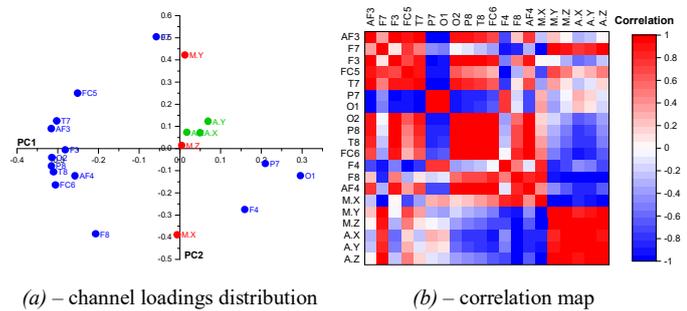

*(a)* – channel loadings distribution    *(b)* – correlation map

Figure 5. Walking: representation of main PCA analysis results

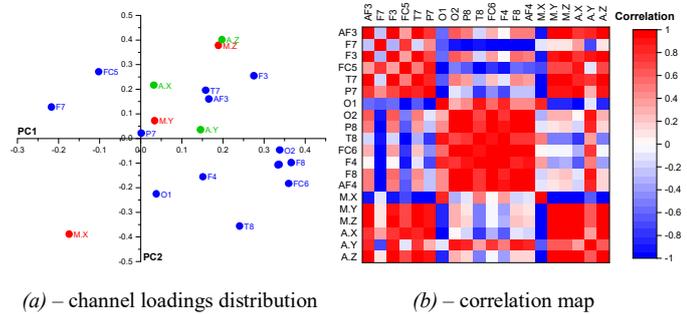

*(a)* – channel loadings distribution    *(b)* – correlation map

Figure 6. Head shaking: representation of main PCA analysis results

As a result of analysing the correlation map for the head shaking (see Fig. 6) we make the following statements:

- strong correlation within all channels of the movement sensors, except the magnetometer channel M.X – it has anti-correlation;
- EEG signals from the left cerebral hemisphere, (electrodes AF3, F7, F3, FC5, T7, P7, except O1) correlate with each other, and the same relation exists for the right hemisphere (see electrodes O2, P8, T8, FC6, F4, F8 and AF4);
- most of the EEG channels of the left cerebral hemisphere have anti-correlation or no correlation with the right hemisphere and vice versa;
- strong correlation between EEG signals from the left cerebral hemisphere and the movement sensors (except the EEG channel F7);
- anti- and no correlation between EEG signals from the right hemisphere and movement sensors (except for accelerometer channel A.Y).
- F7 indicates moderate anti-correlation with the left hemisphere and no correlation with the motion sensors;
- O1 shows strong anti-correlation with the left cerebral hemisphere.

For the walking experiment (see Fig. 5) relations between left, right cerebral hemispheres and motion sensors in general are the same as for the head movement test. In addition to this, channels P7 and O1 have very strong anti-correlation with all EEG signals, however, they correlate with themselves. F7 and F4 also indicate moderate anti- and no correlation with other EEG channels. Here F7 shows strong positive relation with the motion sensors. Correlation of the motion sensors with the left cerebral hemisphere is less pronounced and anti-correlation with the left hemisphere was becoming much stronger.

Such relations are also seen in loading plots, but less distinctive, particularly for the loadings near zero.

As a result we have concluded that:

- Poor contact quality of EEG channels O1 and P7 can certainly be detected using PCA; the known contact quality of the headset electrodes confirm the conclusion about these poor contacts.
- Our previous assumption about poor contact quality of the channels F7 and F4 for the walking supported by theirs relatively low amplitude derived from the visual analysis of the EEG recording is confirmed by PCA but is not confirmed by the quality information exported from the headset. This means that PCA may additionally detect non-related signal differences to the quality.
- The assumption about poor contact quality of the channels F8, F3 and AF3 is not confirmed by both PCA and quality information exported from the headset. This means that these electrodes virtually have a firmly contact and their low amplitude may be explained by their forehead position.
- Data obtained during the EEG recording from the accelerometer and magnetometer cannot be used to detect EEG electrodes with the poor contact quality. However this can be achieved by the calculating the correlation between EEG channels only.

Eyes blinking has scanty motion activity and thus very weak signals from the motion detectors (see Fig. 4-*(c)*).

Thereby there is no sense to investigate the correlation between EEG and motion detectors signals in this case. For the eyes blinking as well for sitting, lying and other steady activities another kind of poor signal detection is proposed and investigated in the next section.

IV. Assessment of the Industrial Power Network Interference

Another possible source of information about EEG channels with poor contact quality is the analysis of signals induced by the industrial power network. Such interference is almost always present. Equal parameters of different electrodes and amplifiers should provide a nearly equal level of the interference signal at 50 Hz or 60 Hz. Larger energy of this spectral component on some electrode may appear due to bad contact of the electrode with the skin, different amplifier gain or due to the EEG signal itself. Such information may be applied to determine EEG channel with the poor contact to the head skin or with the device terminal or amplifier gain reduction.

We calculated the Power Spectral Density (PSD) of EEG channels using Welch periodogram with Hanning window that provides a spectral resolution of 1 Hz. Fig. 7 displays values of 50 Hz PSD component normalized on the maximum value between channels for walking, head motion and eyes blinking.

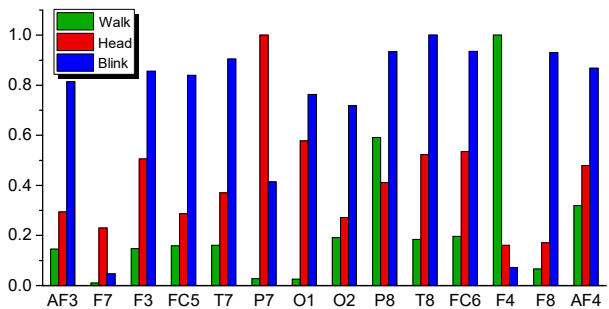

Figure 7. Normalized level of 50 Hz spectral component in the EEG channels

The electrode with a poor contact with the skin can be detected by assessment of the 50 Hz component in two cases: the affected electrode picks up the interference signal with a larger amplitude than the brainwaves and when the electrode has no contact and as a result 50 Hz component will be low. In this manner we supposed that 25% of EEG channels have a poor contact. Our criteria states that this channel should have the highest or the lowest 50 Hz component. Thus we selected one channel with the highest impact and three with the lowest one for each of the three activities.

As a result for the walking these channels are F4, F7, O1, P7; for the head shaking – P7, F4, F8, F7; for blinking – T8, F7, F4, P7. Certainly known channel P7 with a poor contact is presented for all three activities. Channels F4 and F7 also are presented in the each case and additionally they are highlighted by PCA, but the headset did not indicate any quality decrease during the recording. This means that these channels actually have an unstable contact. Certainly known channel with a poor contact O1 is detected only for the walking case. In the other cases for the channels F8 and T8 the detection reason was probably artefacts or another influences.

In general, the results of the described approach correspond to the previously implemented PCA and known information about the failed channels, however a reliable detection of the failed channels is not always possible.

V. Conclusion

Results of this paper demonstrate that there is no significant dependence between electroencephalography signals from channels with bad contacts and signals from motion sensors (accelerometer and magnetometer). Calculation and analysis of the correlation between EEG channels is sufficient to detect failed channels. Analysis of 50 Hz interference from industrial power network in different electroencephalographic channels demonstrates a large difference of this component for failed channels and degraded signals. We consider that the use of additional sensors (such as motion skin impedance meter, electrical muscle activity detector, oxygenation level, etc.) in combination with electroencephalography signals can increase the resilience of the wireless EEG sampling by early detection of failed channels and degraded signals and their further correction. Automated correction of such sampling errors is possible in most cases but additional investigation of this problem is required.

References

[1] R. Ross, R. Graubart, D. Bodeau and R. Mcquaid, "Systems Security Engineering. Cyber Resiliency Considerations for the Engineering of Trustworthy Secure Systems," NIST Special Publication 800-160, vol. 2, 2018.

[2] R. Levalle and S. Nof, "Resilience in supply networks: Definition, dimensions, and levels," Annual Reviews in Control, vol. 43, 2017, pp. 224-236.

[3] Y. Fang, N. Pedroni and E. Zio, "Resilience-based component importance measures for critical infrastructure network systems," IEEE Transactions on Reliability, vol. 65, 2016, pp. 502-512.

[4] Z. Dyka, E. Vogel, I. Kabin, M. Aftowicz, D. Klann, and P. Langendörfer, "Resilience more than the Sum of Security and Dependability: Cognition is what makes the Difference," 8th Mediterranean Conference on Embedded Computing (MECO), 2019, pp. 1-3.

[5] E. Mazomenos, D. Biswas, A. Cranny, A. Rajan, K. Maharatna, J. Achner, .St. Ortmann and P. Langendörfer, "Detecting elementary arm movements by tracking upper limb joint angles with MARG sensors," IEEE journal of biomedical and health informatics, vol. 20 (4), 2016, pp. 1088-1099.

[6] A. Dabbaghian, T. Yousefi, P. Shafia, S. Fatmi and H Kassiri, "A 9.2-Gram Fully-Flexible Wireless Dry-Electrode Headband for Non-Contact Artifact-Resilient EEG Monitoring and Programmable Diagnostics," Proc. IEEE International Symposium on Circuits and Systems (ISCAS), 2019, pp. 1-5.

[7] O. Sudakov, G. Kriukova, R. Natarov, V. Gaidar, O. Maximyuk, S. Radchenko and D. Isaev, "Distributed system for sampling and analysis of electroencephalograms," Proc. IEEE International Conference on Intelligent Data Acquisition and Advanced Computing Systems: Technology and Applications (IDAACS), vol. 1, 2017, pp. 306-310.

[8] V. O. Gaidar and O. O. Sudakov, "Archiving and analysis of electroencephalograms in Ukrainian grid: the first application," IEEE 8th International Conference on Intelligent Data Acquisition and

Advanced Computing Systems: Technology and Applications (IDAACS), vol. 2, 2017, pp. 961-965.
[9] O. Sudakov, M. Kononov, Ie. Sliusar and A. Salnikov, "User Clients for Working with Medical Images in Ukrainian Grid Infrastructure", Proc. International Conference on Intelligent Data Acquisition and Advanced Computing Systems: Technology and Applications (IDAACS), 2013, pp. 705-710.
[10] H. Feldwisch, B. Schelter, M. Jachan, J. Nawrath, J. Timmer, and A. Schulze, "Joining the benefits: combining epileptic seizure prediction methods," Epilepsia, vol. 51(8), 2010, pp. 1598–1606.
[11] A. Aarabi, R. Grebe and F. Wallois, "A multistage knowledge-based system for EEG seizure detection in newborn infants. Clinical Neurophysiology," vol. 118(12), 2007, pp. 2781–2797.
[12] R. R. Fletcher, S. Tam, O. Omojola, R. Redemske and J. Kwan, "Wearable sensor platform and mobile application for use in cognitive behavioral therapy for drug addiction and PTSD," Proc. Annual International Conference of the IEEE Engineering in Medicine and Biology Society, 2011, pp. 1802–1805.
[13] S. Movassaghi, M. Abolhasan, J. Lipman, D. Smith and A. Jamalipour, "Wireless body area networks: A survey," IEEE Communications surveys & tutorials, vol. 16(3), 2014, pp. 1658–1686.
[14] R. Schmidt, T. Norgall, J. Mörsdorf, J. Bernhard and T. von der Grün, "Body Area Network BAN - a key infrastructure element for patient-centered medical applications," J. Biomed Tech, vol. 47, no. 1, 2002, pp. 365–368.
[15] International League Against Epilepsy, https://www.ilae.org.
[16] Shop Fitbit | Fitness Trackers, Smartwatches and More, https://www.fitbit.com/au/store.
[17] M. Manford, Recent advances in epilepsy, Journal of Neurology, vol. 264, 2017, pp. 1811–1824.
[18] A. Catala, M. Grandgeorge, J. L. Schaff, H. Cousillas, M. Hausberger, and J. Cattet, "Dogs demonstrate the existence of an epileptic seizure odour in humans," Scientific reports, vol. 9(1), 2019, 4103.
[19] E. Musk and Neuralink, "An Integrated Brain-Machine Interface Platform With Thousands of Channels", J Med Internet Res, vol. 21(10), 2019, e16194.
[20] T. Ball, M. Kern, I. Mutschler, A. Aertsen, and A. Schulze-Bonhage, "Signal quality of simultaneously recorded invasive and non-invasive EEG," Neuroimage, vol. 46, 2009, pp. 708–716.
[21] X. Xing, Y. Wang, W. Pei, et al. "A High-Speed SSVEP-Based BCI Using Dry EEG Electrodes," Scientific Reports, vol. 8, 2018, 14708.
[22] Featured Products – Rhythmlink, https://rhythmlink.com/products.
[23] DSI Series Dry EEG Headsets - Wearable Sensing https://wearablesensing.com.
[24] V. D. Bram, J. Wouters and M. Moonen, "Optimal electrode selection for multi-channel electroencephalogram based detection of auditory steady-state responses," The Journal of the Acoustical Society of America vol. 126.1, 2009, pp. 254–268.
[25] D. Arnaud, T. Sejnowski and S. Makeig, "Enhanced detection of artifacts in EEG data using higher-order statistics and independent component analysis," Neuroimage, vol. 34.4, 2007, pp. 1443–1449.
[26] N. Hugh, R. Whelan and R. B. Reilly, "FASTER: fully automated statistical thresholding for EEG artifact rejection," Journal of neuroscience methods, vol. 192.1, 2010, pp. 152–162.
[27] EEG Headsets Comparison Chart - Features and Technical Specs, https://www.emotiv.com/comparison.
[28] K. Stamps and Y. Hamam, "Towards inexpensive BCI control for wheelchair navigation in the enabled environment - a hardware survey," Proc. of the 2010 International Conference on Brain Informatics, 2010, pp. 336–345.
[29] EDF files - EmotivPRO v2.0, https://emotiv.gitbook.io/emotivpro-v2-0/edf-files.
[30] J. W. Eaton, D. Bateman, S. Hauberg, and R. Wehbring, "GNU Octave version 4.0.0 manual: a high-level interactive language for numerical computations," 2015.